\documentclass[12pt,a4paper]{article}
\title{Coherent States and the Reconstruction of Pure Spin States}
\author{Jean-Pierre Amiet and Stefan Weigert \\
Institut de Physique, Universit\'e de Neuch\^atel\\
Rue A.-L. Breguet 1, CH-2000 Neuch\^atel, Switzerland\\
\tt stefan.weigert@iph.unine.ch}
\date{April 1999}
\newcommand\be{\begin{equation}}
\newcommand\ee{\end{equation}}
\newcommand\bea{\begin{eqnarray}}
\newcommand\eea{\end{eqnarray}}
\newcommand\ket[1]{|#1\rangle}
\newcommand\bra[1]{\langle #1|}

\begin{document}
\maketitle
\begin{abstract}
Coherent states provide an appealing method to reconstruct efficiently the pure state 
of a quantum mechanical spin $s$. A Stern-Gerlach apparatus is used to measure $(4s+1)$ expectations of projection operators on appropriate coherent states in the unknown state. These measurements are compatible with a finite number of states which can be distinguished, in the generic case, by measuring one more probability. In addition, the present technique shows that the zeroes of a Husimi distribution do have an {\em operational} meaning: they can be identified directly by measurements with a Stern-Gerlach apparatus. This result comes down to saying that it is possible to resolve experimentally structures in quantum phase-space which are smaller than $\hbar$.        
\end{abstract}
To determine the unknown state of a quantum system poses a problem which is interesting from a theoretical, an experimental, and a conceptual point of view. To set up  a successful approach to state reconstruction (see \cite{raymer97} and \cite{leonhardt97} for reviews) means to identify a collection of observables, a {\em quorum}, such that their expectation values contain all the information about the state. As for other inverse problems, ``trial and error'' is the typical approach since no systematic method is known to find an exhaustive quorum. The theoretician might focus on the search for a quorum easily realizable in the laboratory providing certainly a restriction and possibly an inspiration. On the experimental side, high standards 
in the preparation of individual quantum systems are needed since the measurement of 
expectation values requires reliable input states. By now, these standards have been met for quantum systems such as an electromagnetic wave \cite{smithey+93}, vibrating molecules \cite{dunn+95}, ions caught in a trap \cite{leibfried+96}, and atoms moving freely in space after scattering from a double slit \cite{kurtsiefer+97}. Conceptually, 
methods of state reconstruction offer exciting new perspectives since they allow one to faithfully represent quantum mechanics in unconvential ways eliminating the wave function (or the statistical operator) in favour of positive probability distributions \cite{mancini+97,manko+99} or expectation values \cite{weigert99/1}. In a similar vein, a quorum turns out to be an all-embracing source of information allowing one even to determine the expectation values of operators for which no measuring apparatus is available \cite{weigert98/1}.   

A brief summary of early work including the reconstruction of both pure and mixed spin states can be found in \cite{weigert92}. Inspired by tomographic methods to reconstruct the density matrix of a particle, an entirely different approach has been worked out in \cite{leonhardt95}, and it has been adapted in \cite{walser+96} to 
determine a single quantized cavity mode. Coherent states are used as a tool for state reconstruction of spin density matrices in \cite{amiet+99/2}, an approach which turns out to be relatively straightforward. 

Taking into account the extra information that the state to be reconstructed is 
{\em pure}---thus characterized by considerably less parameters than a mixed state---does not automatically simplify the problem. On the contrary, to eliminate the redundancy from the quorum
makes it difficult to set up constructive methods of state determination \cite{amiet+99/1}. 
However, as shown in the following, coherent states offer a new and simple approach to determine a pure spin state by measurements with a Stern-Gerlach apparatus. 
    
The pure states $\ket {\psi }$ of a spin of magnitude $s$ belong to a Hilbert space  $\mathcal{H}_s$ of complex dimension $(2s+1)$, which carries an irreducible representation of the group $SU(2)$. The components of the spin operator $\widehat{\bf S} \equiv \hbar 
\widehat{\bf s}$ satisfy the commutation relations $[{\hat s}_x,{\hat s}_y]= i{\hat s}_z, \ldots$, and they generate rotations about the corresponding axes. The standard basis of the space $\mathcal{H}_s$ is given by the eigenvectors of the $z$ component of the spin, $ {\hat s}_z = {\bf n}_z \cdot {\hat {\bf s}} $, and they are denoted by $\ket{\mu, {\bf n}_z}, -s\leq\mu\leq s$. The phases of the states are fixed by the transformation under the anti-unitary time reversal operator $\widehat{T}$:  $\widehat{T}\ket{\mu,{\bf n}_z}=(-1)^{s-\mu}\ket{-\mu,{\bf n}_z}$, and the ladder operators ${\hat s}_{\pm}= {\hat s}_x {\pm} i {\hat s}_y$ act as usual in this basis:
\begin{equation}
{{\hat s}}_ {\pm} \ket{\mu,{\bf n}_z}
            = \sqrt{s(s+1)-\mu(\mu\pm 1)} \ket{\mu \pm 1,{\bf n}_z} \, .
\label{splusaction}
\end{equation}
With respect to the $z$ basis one has
\be
\ket {\psi } = \sum_{\mu=-s}^s \psi_\mu \ket{\mu, {\bf n}_z} \, ,
\label{expandinz}
\ee
and the ray $\ket {\psi }$ is seen to be determined by $4s$ real parameters since it is normalized to one and its overall phase is irrelevant. For the present purpose, however, a  characterization different from (\ref{expandinz}) will be more convenient. 

Consider the eigenstates of the operator ${\bf n} \cdot {\hat {\bf s}}$, 
\be
{\bf n} \cdot {\hat {\bf s}} \, \ket{ \mu, {\bf n}} = \mu \, \ket{ \mu,{\bf n}} \, , 
\qquad -s \leq \mu \leq s \, ,
\label{Sneigest}
\ee
where the unit vector ${\bf n}= ( \sin \vartheta \cos \varphi,$ $\sin \vartheta \sin \varphi, \cos \vartheta)$, $0 \leq \vartheta \leq \pi, 0 \leq \varphi < 2\pi$, defines a direction in space.
The probability $p_s ({\bf n})$ to measure the value $s$ with a Stern-Gerlach apparatus oriented along ${\bf n}$ equals 
\be
p_s ({\bf n}) = \bra{s, {\bf n} } \, \psi \, \rangle \bra{ \psi } s, {\bf n} \rangle 
              = \bra { \psi } \, z \, \rangle \bra { z } \psi \rangle \, ,
\label{prob}
\ee
given thus by the expectation value of the operator $  \ket{ z } \bra { z }$ projecting 
on a coherent state \cite{arecchi+72},
\be
\ket{ z } \equiv \ket{s, {\bf n} }  
    = \exp [ -i \, \vartheta \, {\bf m}(\varphi) \cdot {\hat {\bf s}} \, ] \, \ket{s,{\bf n}_z} \, ,
\label{defcs}
\ee
where ${\bf m}(\varphi) = (- \sin \varphi,\cos\varphi,0)$. In other words, the coherent state $\ket{ z } $ is obtained from rotating the state $\ket{s,{\bf n}_z}$ about the axis ${\bf m}(\varphi)$ in the $xy$ plane by an angle $\vartheta$. It is convenient to combine $(\vartheta,\varphi)$ into a single complex variable, $z= \tan(\vartheta/2)\exp [ i \varphi ]$. This provides a stereographic projection of the surface of the sphere to the complex plane. In terms of $z$, a coherent state has the expansion \cite{amiet+91}
\be
\ket{ z }  
     = \frac{1}{(1+|z|^2)^s}\sum_{k= 0}^{2s}
           \left(
                  \begin{array}{c}
                      2s \\ 
                       k 
                  \end{array}
           \right)^{1/2} 
z^{k}\ket{s-k, {\bf n}_z} \, .
\label{expandcs}
\ee
Project the state $\ket{ \psi }$ in (\ref{expandinz}) onto the coherent state $\ket{ z }$:
\be
\bra{\psi} z \rangle  
     = \frac{1}{(1+|z|^2)^s}\sum_{k= 0}^{2s}
                 \left(
                  \begin{array}{c}
                      2s \\ 
                       k 
                  \end{array}
           \right)^{1/2} 
\psi^*_{s-k} \, z^{k} 
=  \frac{\psi^*_{-s}}{(1+|z|^2)^s} \prod_{n = 1}^{2s} \left( z - z^0_n \right) \, ,
\label{analyticcoeff}
\ee
providing thus a polynomial of degree $2s$ in $z$ which has been factorized using its $2s$ complex roots $z^0_n$,
some of which may actually coincide. To specify their location in the complex plane requires $4s$ real parameters; therefore, they represent another way to characterize the pure state $\ket{ \psi }$. It has been assumed in (\ref{analyticcoeff}) that the coefficient $\psi^*_{-s}$ is different from zero which is true for almost all states. This ``representation'' of a spin state by $2s$ points in the complex plane \cite{leboeuf+90} has been used, for example, to study the quantum mechanical time evolution of spin systems possessing a classically non-integrable counterpart \cite{leboeuf91}. Since the location of the zeroes $z^0_n$ is easy to visualize, one has a convenient tool to study the quantum dynamics of the system through the paths traced out by the zeroes $z^0_n$ in the course of time.  

The {\em Husimi} distribution \cite{husimi40}, given by the 
squared modulus of (\ref{analyticcoeff}),
\be
P(z^*,z)= | \bra{\psi} z \rangle |^2 
                =  \frac{| \psi_{-s} |^{2}}{(1+|z|^2)^{2s}} 
                       \prod_{n = 1}^{2s} {\left| z - z^0_n \right|}^2 \, .
\label{husimi}
\ee
is a non-negative function in the complex plane or, equivalently, on the classical phase space of the spin. Obviously, the Husimi distribution has the same zeroes as does the function $\bra{\psi} z \rangle $. This observation allows one to implement a first  method to reconstruct the spin state $\ket{ \psi }$, a method which will turn out to be important from a conceptual point of view. According to Eq.\ (\ref{prob}), one just needs to finds those $2s$ orientations ${\bf n}_n$, $n=1,\ldots,2s$, of the Stern-Gerlach apparatus for which the probability to detect the value $s$ vanishes. Experimentally, this means to perform a two-parameter search which may be tedious but, in principle, is possible. Once the directions ${\bf n}_n$ and hence the zeroes $z^0_n$ have been determined, the state $\ket{ \psi }$ is known: multiply out the factors of the product in (\ref{analyticcoeff}) and you can read off the ratios $\psi^*_{s-k}/\psi^*_{-s}$. 

Before turning to a systematic approach to reconstruct the state $\ket{ \psi }$, it should be emphasized that, to the best of our knowledge, the {\em operational meaning} of the zeroes $z_n^0$ exploited here has not been pointed out earlier. On the contrary, doubts have been cast on the physical meaning of the zeroes of the function $P(z^*,z)$ since one thought it impossible to access them experimentally, due to the coarse structure of the quantum phase space. The uncertainty principle for canonically conjugate variables such as position and momentum was assumed to  exclude the resolution of structures in phase space within an area of order $\hbar$. For the spin, however, the vanishing of the Husimi distribution at a {\em point} in phase space can be tested in principle by means of a Stern-Gerlach apparatus. For a particle, 
a similar resoning is expected to apply. If one were able to measure the probability to find the system at hand in whatever coherent state, point-like structures in phase space such as the zeroes of its Husimi distribution would immediately be endowed with physical meaning.  

Here is a {\em systematic} approach to reconstruct a pure spin state based on similar ideas. It is a two-step procedure: first, the probabilities for the value $s$ to occur are measured with a Stern-Gerlach apparatus oriented along $(4s+1)$ 
prescribed directions in space. These experimental data are compatible with $2^{2s}$ pure states which can be written down explicitly. Second, by an additional set of mesurements along $2s$ directions, which are {\em calculated} on the basis of the first series of experiments, one can determine the spin state unambigously. Alternatively, the measurement of one {\em single} probability is sufficient in most cases. 

Step I: On the real axis, $z \equiv x \in I\!\!\!R$, the Husimi distribution turns into a real polynomial of degree $4s$ multiplied with a nonzero factor:
\be
P( x, x )  
                =  \frac{| \psi_{-s} |^{2}}{(1+x^2)^{2s}} 
                       \prod_{n = 1}^{2s} \left( x^2 - 2 u_n x + | z^0_n |^2 \right)
                =  \frac{1}{(1+x^2)^{2s}} \sum_{\lambda =0}^{4s} {\sf c}_\lambda x^\lambda \, ,
\label{realhusimi}
\ee
where $z^0_n = u_n+iv_n$, and the expansion coefficients are functions of the zeroes:
${\sf c}_\lambda = {\sf c}_\lambda(z^0_1,z^0_2,\ldots,z^0_{2s})$. Select now $(4s+1)$ {\em distinct} points $x_\nu $ on the positve real axis,
\be
x_\nu \equiv \tan \frac{\vartheta_\nu}{2} > 0 \, , 
          \qquad x_\nu \neq x_{\nu'} \, , 
          \qquad  0 \leq \nu, \nu' \leq 4s \, ,
\label{pointsplane}
\ee
which correspond to $(4s+1)$ points ${\bf n}_\nu$ of the sphere on the great (semi-) circle with coordinates $\varphi_n =0$ and $\vartheta_n$ following from (\ref{pointsplane}). To fix ideas, one might pick 
\be
x_\nu = \frac{\nu + 1}{4s+1 - \nu} \Rightarrow
\cos \vartheta_\nu 
                = \frac{2s -\nu}{2s+1} \, ,  \qquad \nu = 0,1, \ldots, 4s \, ,
\label{pointssphere}
\ee
providing directions ${\bf n}_\nu$ with equidistant projections on the axis ${\bf n}_z$.  
Measure now the probabilities to obtain the value $s$ along the directions ${\bf n}_\nu$; this results in $(4s+1)$ numbers 
\be
 P (x_\nu,x_\nu) \equiv {\sf p}_\nu 
         =  \sum_{\lambda =0}^{4s}  {\sf M}_{\nu\lambda} {\sf c}_\lambda \, , \qquad 
            {\sf M}_{\nu\lambda} = \frac{(x_\nu)^\lambda}{(1+x_\nu^2)^{2s}} \, ,
\label{firsteq}
\ee
with a real matrix ${\sf M}_{\nu\lambda}$ of dimension $(4s+1) \times (4s+1)$, the entries of which are all between $0$ and $1$. Eq. (\ref{firsteq}) can be solved for the coefficients ${\sf c} = ({\sf c}_0, {\sf c}_1, \ldots, {\sf c}_{4s})^T$ according to 
\be
{\sf c} = {\sf M}^{-1} \,  {\sf p} \, ,
\label{secondeq}
\ee
since the matrix ${\sf M}$, being of type Vandermonde \cite{gradshteyn+80}, has a nonzero determinant,
\bea 
\det {\sf M} & = & 
\left( \prod_{\nu = 0}^{4s} \frac{1}{(1+x_\nu^2)^{2s}} \right) 
\left|
\begin{tabular}{cccc}
$1$      &  $x_0$   & $\cdots$ & $(x_{0})^{4s}$ \\
$\vdots$ & {}       & {}       & $\vdots$      \\
$1$      & $x_{4s}$ & $\cdots$ & $(x_{4s})^{4s}$ 
\end{tabular}
    \right| \nonumber \\
& = & \left( \prod_{\nu = 0}^{4s} \frac{1}{(1+x_\nu^2)^{2s}} \right) 
\prod_{0\leq \nu < \nu' \leq 4s} (x_{\nu'}-x_\nu) \, ,
\label{fund-matrix}
\eea
since $x_\nu \neq x_{\nu'}$, $\nu \neq \nu'$, according to (\ref{pointsplane}).
An explicit expression of the inverse of a Vandermonde matrix is given in {\cite{newton+68}.

Knowing the coefficients ${\sf c}_\lambda$ obtained with (\ref{secondeq}), the function $P (x,x)$ is determined through the sum in Eq.\ (\ref{realhusimi}), and one can work out its $2s$ factors $( x^2 - 2 u_n x + | z^0_n |^2 )$. This, however, is {\em not} sufficient to determine
the state $\ket{ \psi }$: generically, there are $2^{2s}$ distinct states giving rise to  the {\em same} expansion coefficients ${\sf c}$. To see this consider the state $\ket{ \psi (j) }$, $j \in 
(1,\ldots,2s)$ with 
\be
\bra{\psi (j)} z \rangle  
     =  \frac{\psi^*_{-s}(j)}{(1+|z|^2)^s} \, \left( z - (z^0_j)^* \right) \, \prod_{n = 1, n \neq j}^{2s} 
        \left( z - z^0_n \right)  \, ,
\label{analyticcoeffj}
\ee
having thus the same zeroes as $\ket{ \psi }$ except for $z^0_j$ which has been replaced by its complex conjugate $(z^0_j)^*$. The Husimi distributions $P (z^*,z) $ and $ P_j (z^*,z) = 
| \bra{ \psi (j) } z \rangle|^2$ differ from each other for general $z$ but for {\em real} arguments (cf.\ Eq.\ (\ref{realhusimi})) they coincide:
\be
P (x,x) = P_j (x,x) \, , \qquad j \in (1, \ldots, 2s) \, ,
\label{real=eq}
\ee
since the information about the {\em sign} of the imaginary part of the $j$-th zero has 
dropped out. This is also obvious from looking more closely at the factors $( x^2 - 2 u_n x + | z^0_n |^2)$ in Eq.\ (\ref{realhusimi}): they allow one to calculate unambiguously the real parts $u_n$ of the zeroes but only the {\em modulus} of their imaginary parts, $| v_n |^2 = | z^0_n |^2 - | u_n |^2$. Thus, a twofold ambiguity is present for each of the $2s$ zeroes of $\ket{\psi}$, giving rise to a total of $2^{2s}$ states\footnote{This ambiguity happens to be smaller in exceptional cases: one or more of the zeroes might be  real, or some zeroes might coincide.} compatible with the data ${\sf p}$. It is straightforward to remove the ambiguity by a second series of measurements.   

Step II: It remains to find out whether a zero is located at $z^0_n $ or at $(z^0_n)^* $, $n=1, \ldots, 2s$. To do this, one orients the Stern-Gerlach apparatus along the axis ${\bf n}_n$ or 
along ${\bf n}_{n^*}$ associated with the point $z^0_n $ or its complex conjugate, respectively. If the probability to measure the value $s$ along ${\bf n}_n$ does {\em not} vanish, the zero in question is located at ${\bf n}_{n^*}$. If it vanishes, the zero is determined by the actual orientation of the apparatus. Repeating this procedure for the $2s$ pairs of points, one is able to remove the ambiguity completely. 

There is a possibility---working in almost all cases---to reduce the measurements required for  Step II to the determination of a {\em single} probability.  The $2^{2s}$ states compatible with the data $\sf p$ are known explicitly. Now {\em calculate} the values of the probability to measure the value $s$ along one {\em randomly} chosen direction ${\bf n}'$ in space  which does not coincide with any of the putative zeroes. If the $2^{2s}$ calculated values all differ from each other, it is sufficient to perform the measurement along ${\bf n}'$ in order to identify the correct state. If, however, two or more probabilities with respect to the axis ${\bf n}'$ coincide, a slight change 
${\bf n}'\to {\bf n}''$ usually will be sufficient to remove the degeneracy. Unfortunately, it is difficult to actually determine a particular spatial direction such that the probabilities are {\em guaranteed} 
to have different values---although this is the generic case to happen.
 
When carrying out the first proposal one measures a total of $(4s+1) + 2s = (6s+1)$ real parameters in order to determine the state $\ket{\psi}$. The second method requires only $(4s+1) +1 = 2(2s+1)$ measurements which exceeds the number of free parameters by two 
only. It is essential, however, that both cases give results {\em linear} in $s$ providing thus a substantial reduction compared to the optimal method ($\propto s^2$) for a mixed state.       

Points $x_\nu$ on the positive real axis are not the only choice for a successful reconstruction scheme. They might be located, for example, on a half line starting 
at the origin which encloses an angle ${\tilde \varphi} \in (0,2\pi)$ with the real axis. This situation is obtained from the original one by a conformal transformation, $z \to w = z \exp (-i{\tilde \varphi})$, that is, a rigid rotation about the origin by an angle ${\tilde \varphi}$. 

It is also possible to carry out Step I with $(4s+1)$ distinct complex numbers of modulus
one, $\exp[i \varphi_\nu]$, providing directions ${\bf n}_\nu$ on the equator of the sphere.
The elements of the matrix ${\sf M}$ then read
\be
{\sf M}_{\nu\lambda} = \frac{1}{2^{2s}} \left( \exp[i\varphi_\nu] \right)^\lambda \, ,
\label{firsteqequat}
\ee
its determinant being different from zero if no two angles $\varphi_\nu$ coincide. Again, a $2^{2s}$-fold ambiguity remains which is due to the existence of other states with the {\em same} Husimi function now on the unit circle. Consider, for example, the state 
\be
\bra{\psi (j)} z \rangle  
     =  \frac{\psi^*_{-s}(j)\sqrt{|z^0_j|}}{(1+|z|^2)^s} \, 
        \left( z - \frac{1}{(z^0_j)^*} \right) \, 
         \prod_{n = 1, n \neq j}^{2s} \left( z - z^0_n \right)  \, .
\label{analyticcoeffjequat}
\ee
Its Husimi function for the argument $z= \exp[i \varphi ]$ satisfies 
\be
P_j  (e^{-i\varphi},e^{i\varphi}) = P(e^{-i\varphi},e^{i\varphi}) \, ,
\qquad j \in (1, \ldots, 2s) \, ,
\label{real=eqequat}
\ee
so that one cannot distinguish the states $\ket{\psi (j)}$ and $\ket{\psi}$ by looking at the unit circle only. The property (\ref{real=eqequat}) is due to the fact that $P(z^*,z)$ on the unit circle is invariant under the transformation $z \to 1/z^*,$ while the invariance associated with (\ref{real=eq}) is due to a reflection $z \to z^*$ about the real axis.

Further, one can displace the points rigidly, $z \to w = z +{\tilde z}$, ${\tilde z}$ a fixed complex number, subsequently reworking one's way through the derivation given above. A shift of the points on the positive real axis to the left shows once more that one can use also $(4s+1)$ points on a (and therefore: any) great circle. Apply now a shift ${\tilde z}$ to the ensemble of points located on the equator: then, this great circle turns into another circle on the surface of the sphere providing thus another ensemble of directions for the measurements. 

It appears plausible---although no proof is available yet---that Step I can be based on a
set of $(4s+1)$ {\em arbitrary} distinct points in the complex plane. One might speculate that such a generalisation is more effective than the method presented above: possibly {\em no} ambiguity will show up if $(4s+1)$ `random' points in the complex plane are chosen. In other words, the necessity of Step II (to perform additional measurements) could result from the non-generic choice of the points on a {\em line} (or {\em circle}) introducing unintentionally a reflection (or inversion) symmetry. 

Summing up, the method of state reconstruction via coherent-state expectation values can be based on any $(4s+1)$ points in the complex plane located on an arbitrary circle, including its limit of a straight line. In the generic case, one additional measurement is needed to determine the actual pure state of the system. Furthermore, for a pure spin state, a search for the zeroes of its Husimi distribution provides another strategy of state reconstruction since they are seen to be accessible in experiments with a Stern-Gerlach apparatus.                      

\section*{Acknowledgements}
St. W. acknowledges financial support by the {\em Schweizerische Nationalfonds}.  
\end{document}